# Photon Management for Silicon Solar Cells featuring Hole-Selective Molybdenum Oxide Rear Contacts: An Optical Simulation Study

Mohammad Jobayer Hossain[1], Kristopher O. Davis[1, 2]

[1]CREOL, The College of Optics and Photonics, University of Central Florida, Orlando, FL 32816, USA
[2]Department of Material Science and Engineering, University of Central Florida, Orlando, FL 32816, USA

*Abstract* — Passivated, hole-selective contacts play important role in reducing surface recombination by lowering the concentration of electrons in the rear side of a solar cell. However, parasitic optical losses in these contacts can still limit the performance of the cell. In this work, the long wavelength optical losses of silicon solar cells featuring hole-selective molybdenum oxide ($MoO_x$) rear contacts are investigated using optical simulations. The potential of these selective contacts for possible enhancement of photogenerated current density was also investigated for their use with nanostructured dielectric layers.

*Index Terms* — FDTD simulation, grating nanostructure, hole selective contact, photon management, silicon solar cell.

## I. INTRODUCTION

Silicon is still the dominant material in the market for photovoltaic (PV) energy conversion. Although the levelized cost of electricity of PV systems has dropped to incredibly low levels in recent years, there is still room for further improvements in terms of the efficiency, manufacturing cost, and durability of PV cells and modules. To maximize cell efficiency, the optical losses, recombination losses, and resistive losses [1], [2] of solar cells must be minimized.

In traditional *p*-type cell architectures (e.g., Al-BSF, PERC), Shockley-Read-Hall recombination at the metal/semiconductor interfaces of the electrical contacts can be a dominant loss mechanism that lowers the potential efficiency of these cells. Passivated, carrier-selective contacts can help reduce this contact recombination loss, and considerable effort is being put into the development of both electron and hole-selective contacts using various materials, including doped amorphous silicon [3], doped polycrystalline silicon [4]–[6], transition metal oxides [7]–[13], and transition metal nitrides [14].

Molybdenum oxide ($MoO_x$) is one such material that acts as a hole-selective contact when deposited on silicon. The high work function of $MoO_x$ induces band bending in the silicon that can drastically lower the concentration of electrons at the silicon surface. However, the magnitude of this band bending is dependent on the type of passivation material (if any) used between the silicon and the $MoO_x$, the thickness of the $MoO_x$ film, the deposition process used, the metal put in direct contact with $MoO_x$, and any post-deposition annealing. This is due, in large part, to relationship between the work function of $MoO_x$ and the concentration of oxygen vacancies present in the films.

In this work, the optical losses of silicon solar cells featuring hole-selective molybdenum oxide as both a full area rear contact and local rear contacts are investigated. Here, simulations of the photogenerated current ($J_G$) are performed for monocrystalline *p*-type cells featuring a rear $MoO_x$ contact with different contact metals and contact fractions. In addition to 1D thin film stacks, 2D grating structures are considered to evaluate any potential current gains due to scattering and diffraction. Losses in $J_G$ are broken down to identify the amount of parasitic optical absorption in individual layers. These results are compared to the losses for Al-BSF and PERC cells. An illustration of the cell architectures considered in this work is shown in Figure 1. In all cases, the front side of the cells are considered to be the same, with anisotropically textured random pyramids and a standard silicon nitride ($SiN_x$) anti-reflection coating (ARC).

## II. MODELING METHODOLOGY

In this work, two simulation tools are used. (1) *SunSolve* used to perform ray tracing of the optics of the solar cell and all of the thin films present within the device thus quantify $J_G$, front and escape reflectance loss and parasitic optical absorption. (2) *Lumerical* is used to perform finite difference time domain (FDTD) simulations to evaluate potential improvements in light trapping if nanostructures dielectric layers are used. This tool accounts for reflectance, scattering (if any), absorption, and transmission within the substrate itself, and the results of *Lumerical* are fed back into *SunSolve* to calculate the potential gains in $J_G$.

### A. SunSolve Ray Tracing Simulations

For the ray tracing simulations carried out with *SunSolve*, the same front surface was considered with a random upright pyramids and a 75 nm $SiN_x$ ARC. A wafer thickness of 180 µm was also assumed in each case. For each simulation, the maximum achievable $J_G$ is determined based on the different rear surface configurations. These configurations include: an Al-BSF rear surface with an Al-Si eutectic and $p^+$ BSF; a dielectrically passivated rear surfaces featuring 20 nm of aluminum oxide ($Al_2O_3$) followed by a 100 nm $SiN_x$ film.

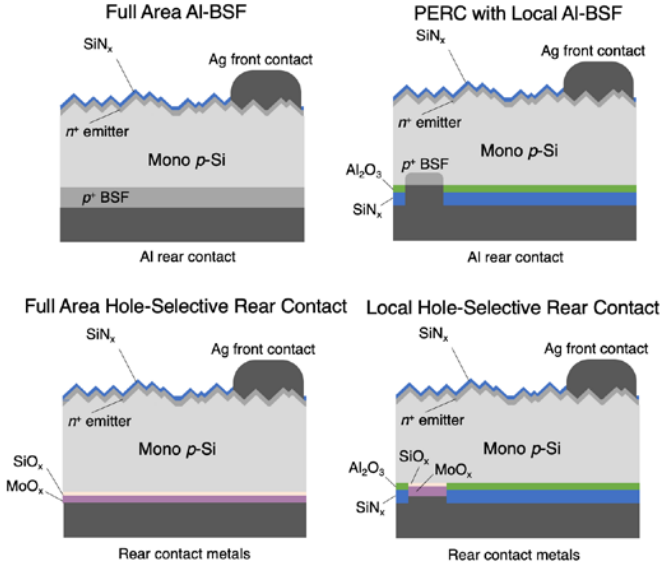

Fig. 1. Illustrations of the cell architectures considered in this optical simulation study.

For the hole-selective MoO$_x$ surfaces, a 2 nm SiO$_x$ layer is assumed followed by a 5 nm MoO$_x$ film. The SiO$_x$ layer is considered since it is known to be present following both thermal evaporation and atomic layer deposition of MoO$_x$. Ultimately, this thin SiO$_x$ has a negligible influence of the rear optics and could be neglected. A number of different rear contact metals were considered. Work by Gregory *et al.* has shown that Ni, capped with Al, forms a more thermally stable, ohmic contact with lower contact resistivity than Al in direct contact with MoO$_x$ [15]. This is due to the higher work function of Ni as well as the lower oxygen affinity. Unfortunately, Ni strongly absorbs the NIR photons that reach the rear surface. The cases where MoO$_x$/Al and MoO$_x$/Ni/Al are both included in the *SunSolve* simulations.

Finally, a Lambertian scattering factor was changed from zero (i.e., purely specular) to one (i.e., purely Lambertian). This factor strongly influences escape reflectance and therefore the overall $J_G$.

*B. Lumerical FDTD Simulations*

FDTD simulations were carried out for a full area hole-selective rear contact and three different local hole-selective rear contact structures demonstrated in Figure 2. The same material stack of Si/SiO$_2$/MoO$_X$/Ni/Al is used for all the contacted regions (Z axis). The surface passivation comes from the periodic grating of rectangular, cylindrical and inverse cylindrical block of Al$_2$O$_3$, spread over the XY plane, as depicted in Figure 2(b), 2(c) and 2(d) respectively. For all the cases, plane polarized light source was used. Periodic boundary conditions were set in both X and Y axis. A lossless silicon is assumed as the medium of the light source. This allows absorption occurring only in the rear contact structures to be quantified.

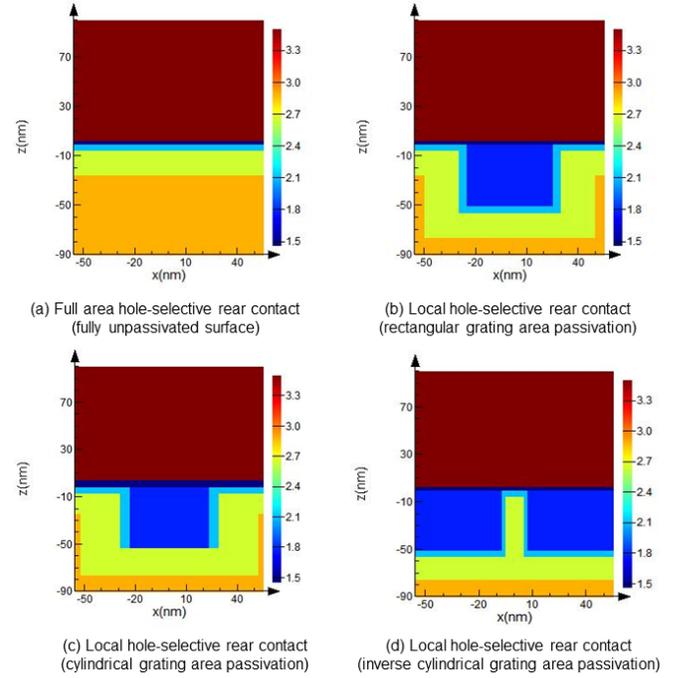

Fig. 2. Hole-selective rear contact structures simulated using *Lumerical*.

All these complex calculations were carried out for the 365 nm to 1280 nm wavelength range. The local area 2D grating structure was optimized specifically for the longer wavelengths. Both normal incidence and other angles of incidence of light on the side were investigated. An angle of incidence of 41.4° is of utmost interest in this study since 76.4% of the total light falls on the rear surface at this angle [16], [17]. The *Lumerical* FDTD simulation provides reflectance, scattering and absorption of the nanostructured contacts which would not otherwise be possible to obtain with the *SunSolve* ray tracing simulations.

III. RESULT AND DISCUSSION

*A. SunSolve Ray Tracing Simulations*

The estimated $J_G$ values calculated using *SunSolve* are shown in Figure 3, for the four different rear surfaces evaluated in this work. The maximum value for each configuration is for the purely Lambertian case with maximum scattering and the minimum value is the purely specular reflection case with no scattering.

As expected, the dielectric passivated Al$_2$O$_3$/SiN$_x$ surface has the maximum expected $J_G$ due to unity internal reflectance [15]. The Al-BSF and MoO$_x$/Ni/Al perform poorly due to significant parasitic optical absorption within the Al-Si eutectic layers for the Al-BSF and in the Ni layer for the MoO$_x$/Ni/Al contact. When considering optical, recombination, and resistive losses, the ideal rear surface would feature local hole-selective MoO$_x$/Ni/Al contact with a dielectric passivated Al$_2$O$_3$/SiN$_x$ surface covering >95% of the rear. Maximum scattering is ideal for this structure, since over 0.8 mA/cm² is gained between the specular case and Lambertian case for a 180 µm wafer. For thinner wafers, the potential current gains due to increased scattering is even higher. Dielectric nanostructures are also considered as a means of maximizing light trapping via scattering and/or diffraction using *Lumerical*.

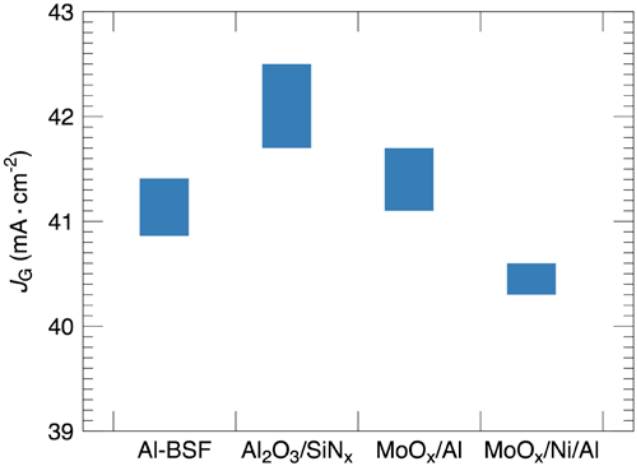

Fig. 3. $J_G$ calculations for different rear surface configurations. The maximum value for each configuration is for the purely Lambertian case with maximum scattering and the minimum value is the purely specular reflection case with no scattering.

*B. Lumerical FDTD Simulations*

The three different local contact structures along with full area contact structure shown in Figure 2, all based on $SiO_2/MoO_X/Ni/Al$ stack, were simulated using *Lumerical*. The full area contact does not have any $Al_2O_3$ between $SiO_2$ and $MoO_X$. For the rectangular and cylindrical grating structures, around 25% of the rear surface has $Al_2O_3$ between $SiO_2$ and $MoO_X$ (50% in X axis and 50% in Y axis). This value is about 99% for the inverse cylindrical grating structures. Figure 4 illustrates the reflectance characteristics of the rear surface featuring these contact structures. The reflectance was calculated by setting a 2D plane wave source normal to the direction of incidence (Z axis) and a 2D detector likewise. Since these are rear surface structures, the reflectance value at longer wavelengths is critical. The inverse cylindrical area passivation structure shown in Figure 2(d) provides the highest value of reflectance, followed by the full area contact, cylindrical passivation and rectangular passivation. Since there is no light being transmitted out of the cell, the reflectance is limited by the parasitic absorption in the contact metals. Smaller value of the absorption loss leads to a higher value of reflectance and $J_G$ as a result.

There are two pathways to the absorption loss in the rear side. The first is the partial propagation of the normal component of the incident light (especially when the $SiO_2$ and $MoO_X$ films are thin) and eventually getting absorbed in the metal.

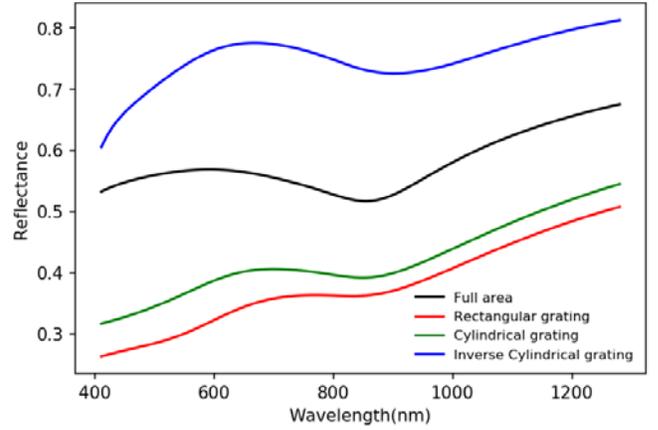

Fig. 4. Reflectance characteristics of the rear surface featuring four different hole-selective rear contact structures

The second one comes from the excitation of the surface plasmon polariton (SPP) due to the presence of any sharp corner. Both the loss mechanisms are present in the other three contact structures. The full area contact, where the second mechanism is absent therefore shows higher reflectance than the rectangular and cylindrical grating structure as illustrated in Figure 4. The cylindrical grating structure, because of its circular cross section (XY plane) is able to avoid some of the SPP excitation. That is why it shows little higher reflectance than the rectangular grating structure. Now, for the inverse cylindrical grating structure, it is able to avoid the first mechanism for the 99% of the surface where Si is not at the vicinity of Ni. Optical losses still occur due to both the mechanisms being present in the rest 1% unpassivated area. The result is a reflectance higher than even the full area unpassivated contact structure.

The above explanations becomes clearer when we take a close look into the parasitic absorption loss profile illustrated in Figure 5. Dissimilar to the reflectance results in Figure 4, which were calculated for the whole surface, the absorption profiles were calculated only for a small area because of the limited computational power available during this study (calculating absorption is way more complicated than the reflection); absorption was calculated for a slice of material going through the center of the nanostructures. Nevertheless, it provides important insights. Since this study is focused on the rear side of the cell, our primary interest is in the longer wavelength photons. We have chosen a long wavelength value, 975 nm for calculating absorption profiles.

The full area contact suffers from significant parasitic absorption, especially in the Ni layer for both normal and 41.4° incidence. Red means more parasitic optical loss and blue means less in the figure. The angled incidence shows less loss through propagation since the normal component of the propagation vector is smaller for the angled incidence. As depicted in Figure 5(b) through 5(d), the $Al_2O_3$ passivated local contact structures are able to avoid part of these losses, the amount being determined by the percentage of the total area passivated. However, it suffers from losses in any sharp corner; this effect arising from the excitation of surface plasmon polariton (SPP) becomes severe for the case of angled incidence in a rectangular local contact structure. Then cylindrical passivated contact structure seems to be able to avoid additional SPP excitation loss coming from the angled incidence because of its round geometry in XY plane; the horizontal component of the light just diffracts around the structure in XY plane. The inverse cylindrical structure still shows absorption loss in the Ni for normal incidence even after passivating most of the regions by $Al_2O_3$, although this is not as severe as the normal incidence in the full area contact. For both the local cylindrical and inverse-cylindrical nanostructures, absorption loss is low for angled incidence.

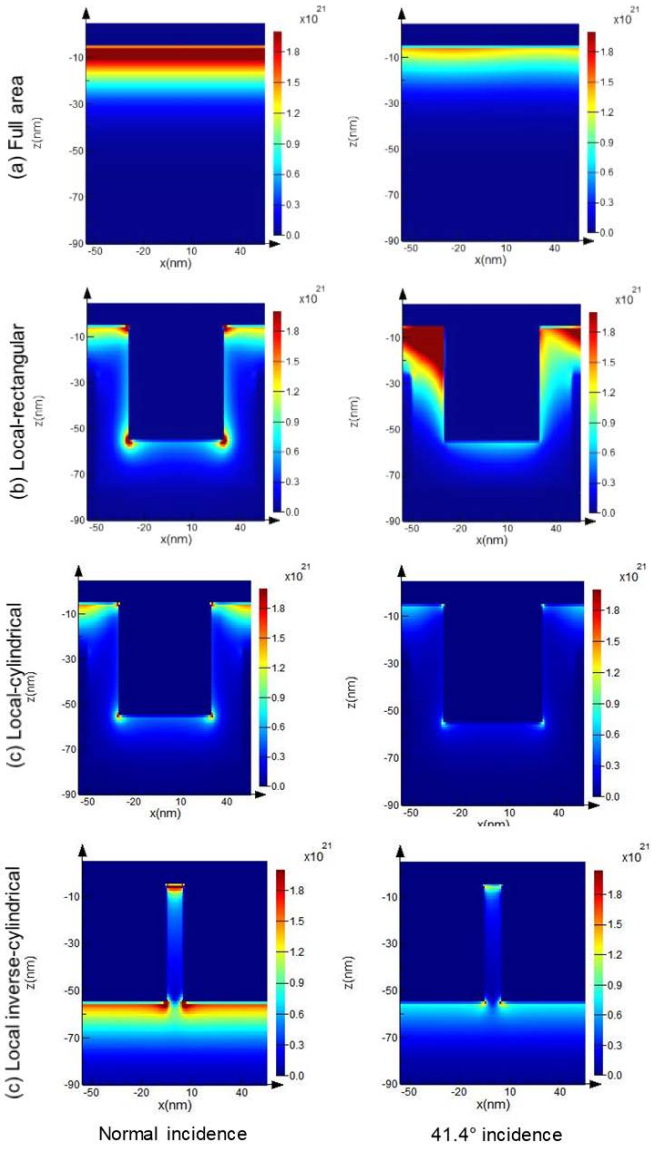

Fig. 5. Losses (W/m³) in different local hole-selective rear contact structures at 975 nm wavelength, both for normal and 41.4° incidence (calculated only for a slab of material going through the center of the structure).

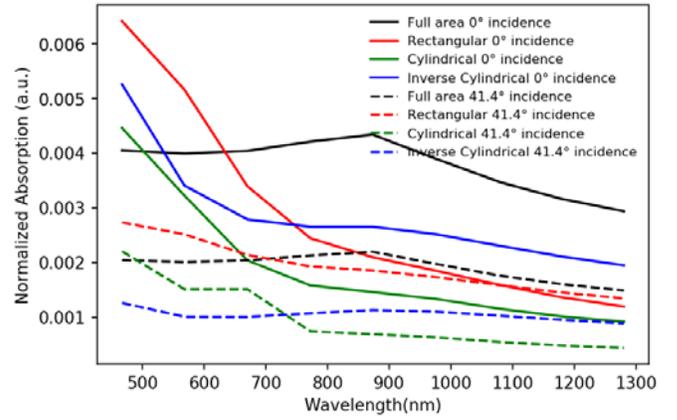

Fig. 6. Absorption loss characteristics of four different hole-selective rear contact structures (calculated only for a slice of material going through the center of the structure).

Figure 6 illustrates the absorption loss characteristics of the structures shown in Figure 5. These values in these graphs again resonates with the loss profiles in Figure 5. Loss is highest in full area contact followed by inverse cylindrical, rectangular and cylindrical structures for normal incidence. For the case of 41.4° incidence, this order is: full area, rectangular, inverse cylindrical and cylindrical structures, at longer wavelengths.

When Figure 5 and 6 provides important insights regarding absorption characteristics, they do not quite agree with the reflectance characteristics in Figure 4 i.e. high absorption should always lead to low reflectance and vice versa.

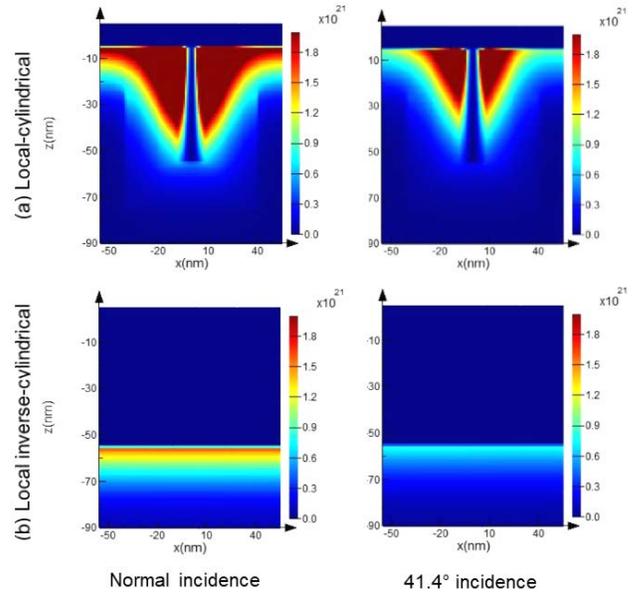

Fig. 7. Losses (W/m³) in two local hole-selective rear contact structures at 975 nm wavelength, both for normal and 41.4° incidence (calculated only for a slab of material at Y=30nm).

The reason lies in the fact that the absorption profiles and the characteristics depicted in Figure 5 and 6 presents the absorption loss scenario for a slab of material going through the middle of the structure (Y=0 nm), when the reflectance graph in Figure 4 is calculated for the whole rear surface (accurate case). To confirm this guess, similar absorption profile was calculated at Y=30 nm and is depicted in figure 7. When the local cylindrical structure in Figure 5 (Y=0 nm) showed low loss, it showed high loss for Y=30 nm. For the inverse cylindrical structure it becomes very similar to the full area case, because we have avoided the structure by choosing Y=30 nm while calculating the absorption profile in Figure 7.

Overall, Figure 4 provides the accurate representation of the reflectance graph, which is dependent on the absorption characteristics of the whole rear surface.

IV. CONCLUSION AND FUTURE WORK

Optical simulations have been carried out to evaluate ways of maximizing the photogenerated current of hole-selective rear contacts. Parasitic optical absorption in the Ni interlayer used is the dominant optical loss mechanism in these structures, similar to the case of Al-BSF cells that features similar absorption in the Al-Si eutectic. As with Al-BSF cells, the use of a dielectrically passivated surface covered the majority of the rear surface can reduce both optical and recombination losses in the $MoO_x$ hole-selective contacts. Comparisons between a full area contact and local contact structures were investigated using both ray tracing and FDTD simulations in this work. Inverse cylindrical passivation nanostructures are found to be the best choice considering high reflection and low absorption it can provide. The structure will be optimized looking into the $J_G$ benefit coming from reflection, scattering and avoiding parasitic absorption losses. Then it will be incorporated in the standard solar cell architecture for better performance.